\documentclass[11pt,preprint]{aastex}

\def\kms{km s$^{-1}$}
\usepackage{lscape}

\begin{document}

\title{ 
No evidence of chemical anomalies in 
the bimodal turnoff cluster NGC~1806 \\ 
in the LMC
\footnote{Based on observations collected at the ESO-VLT under the program 084.D-0933 
and collected with the NASA/ESA HST, obtained at the Space Telescope Science Institute, 
which is operated by AURA, Inc., under NASA contract NAS5-26555.}
}
 
\author{A. Mucciarelli$^{1}$, E. Dalessandro$^{1}$, 
F. R. Ferraro$^{1}$, L. Origlia$^{2}$, B. Lanzoni$^{1}$}

\affil{$^{1}$Dipartimento di Fisica \& Astronomia, Universit\`a 
degli Studi di Bologna, Viale Berti Pichat, 6/2 - 40127
Bologna, ITALY}

\affil{$^{2}$INAF - Osservatorio Astronomico di Bologna, 
via Ranzani, 1 - 40127, Bologna, ITALY}

\begin{abstract} 
We have studied the chemical composition of NGC~1806, a massive, intermediate-age globular cluster 
that shows a double main sequence turnoff.
We analyzed a sample of high-resolution spectra (secured with FLAMES at the Very 
Large Telescope) for 8 giant stars, members of the cluster, finding 
an average iron content of [Fe/H]=--0.60$\pm$0.01 dex and
no evidence of intrinsic star-to-star variations in the abundances of light elements 
(Na, O, Mg, Al).
Also, the ($m_{F814W}$; $m_{F336W}-m_{F814W}$) color-magnitude diagram obtained 
by combining optical and near-UV Hubble Space Telescope photometry exhibits 
a narrow red giant branch, thus ruling out intrinsic variations of C and N abundances in the cluster. 
These findings demonstrate that NGC~1806 does not harbor chemically distinct sub-populations, 
at variance with what was found in old globular clusters. In turn, this indicates  that the double main sequence turnoff 
phenomenon cannot be explained in the context of the self-enrichment processes usually 
invoked to explain the chemical anomalies observed in old globulars. 
Other solutions (i.e., stellar rotation, merging between clusters or 
collisions with giant molecular clouds) should be envisaged to explain this class of globulars.

\end{abstract}

\keywords{stars: abundances --- globular clusters: individual (NGC~1806) --- Magellanic Clouds 
--- techniques: spectroscopic --- techniques: photometric}

\section{Introduction}

Several intermediate-age (1-3 Gyr) globular clusters (GCs) in the Large and Small 
Magellanic Clouds (LMC and SMC) exhibit an extended main 
sequence turnoff (MSTO) in their optical color-magnitude diagrams (CMDs).
The observed spread can appear either as a broad region or as a clear bifurcation, 
while the MS and red giant branch (RGB) appear well defined and narrow in the optical CMDs. 
By using Hubble Space Telescope (HST) photometry \citet{mackey07} first detected 
two distinct MSTOs in the massive, intermediate-age LMC GC NGC~1846 
and in the last few years the list of GCs with anomalous MSTO morphologies has grown significantly
\citep[see e.g. the sample of 16 LMC GCs discussed by][]{milone09}.
According to the observed MSTO morphology, three classes of intermediate-age Magellanic clusters 
can be distinguished:
(1)~GCs with a narrow MSTO, with the massive GC NGC~1978 as prototype
\citep{m07_1978}; 
(2)~GCs with a broad MSTO but no clear bimodality;
(3)~GCs with a bimodal MSTO.
Out of 16 GCs studied by \citet{milone09}, five belong to class (1) and eight 
to class (2). Examples of clusters with bimodal MSTO are
NGC~1751, NGC~1806, NGC~1846 in the LMC \citep{mackey08,milone09,goud09} and NGC~419 in the 
SMC \citep{glatt08}.

In GCs where distinct sequences are observed, population ratios and radial 
distributions have been studied with results that are still debated and controversial.
For example,  \citet{goud09} found that the brightest MSTO stars in NGC~1846 
are more centrally concentrated than the faintest ones.
Instead, in NGC~1783 the faintest MSTO stars are more centrally concentrated \citep{rubele13}.
Finally, at least in the cases of NGC~1806, NGC~1846 and NGC~1751, 
the brightest MSTO appears to be more populated with respect to the faintest one \citep{milone09}.

Spurious effects due to differential reddening, field star contamination or 
photometric uncertainties have been deeply investigated and ruled out \citep[see e.g.][]{mackey08,goud11} 
and different scenarios to explain the observed MSTO have been envisaged. In particular it has been proposed that  
the extended MSTOs could reflect a real age difference, thus suggesting that these 
clusters have undergone prolonged (continuous or bursty) star formation activity 
\citep[see e.g.][]{goud09,milone09,keller11}. Within this framework, the ages should range 
between $\sim$50 and $\sim$300 Myr \citep{milone09}.
An alternative explanation could be that an extended MSTOs are due to the presence of fast rotating 
MS stars, which are predicted to evolve at lower effective temperatures (thus redder colors) 
with respect to non-rotating MS stars. However, conflicting results about the efficiency of this effect have 
been found so far \citep[see e.g.][]{bastian09,girardi11}. 
Finally, through  merging and mass transfer processes, interactive binaries 
can also produce an extension of the MSTO toward bluer colors and brighter magnitudes \citep{yang11}.
However, this  could only partially explain the broad MSTOs \citep[under the strong 
assumption that all the cluster stars are members of binary systems;][]{yang11}, but
this scenario is unable to account for the bimodal MSTO.

In the first case, \citet{goud09} and \citet{keller11} proposed a connection between the double/broad MSTO 
and the self-enrichment processes invoked to explain the chemical anomalies in the light elements 
abundances commonly observed in 
old, massive GCs \citep[see e.g.][and references therein]{gratton12}. 
In the generally accepted scenario 
\citep[see e.g.][]{ventura01,dercole08,decressin10,valcarce11}, the currently old and massive GCs
underwent prolonged  star formation activity in their early stage, over a period
 smaller than $\sim$200-300 Myr. 
Thus, {\sl second generation} stars formed from a gas processed by the CNO cycle and ejected by 
first generation stars,
like asymptotic giant branch (AGB) and/or fast-rotating massive 
stars\footnote{However, 
it is worth recalling that \citet{bastian13a} proposed an alternative mechanism 
to explain the chemical anomalies in GCs without resorting to multiple star formation 
episodes.}. 
Following this scenario, the dual/broad MSTOs would be 
direct evidence of prolonged star formation activity occurring in the early stages of GC formation 
(which is not observable in the CMDs of old GCs because the age spread is negligible with respect 
to the cluster age).
As a consequence, Magellanic GCs with dual/broad MSTO should exhibit intrinsic star-to-star variations 
in the abundances of the elements involved in the complete CNO cycle (C, N, O, Na, Mg, Al).

The aim of this Letter is to test this scenario, 
searching for the expected chemical anomalies in the massive GC 
NGC~1806 \citep[M=1.07$\cdot10^{5}M_{\odot}$,][]{goud11b}, which shows a well-defined bimodal MSTO \citep{mackey08}.

\section{Observations and data reduction}

In order to investigate possible chemical anomalies in the abundances of light elements of NGC~1806, 
we analyzed high-resolution spectra to derive O, Na, Mg and Al abundances, 
and high-resolution photometric data to study C and N abundances.

-- {\it The spectroscopic data set: }
High resolution spectra have been obtained with the multi-object facility FLAMES 
\citep{pasquini} at the Very Large Telescope of the European Southern Observatory in the 
UVES+GIRAFFE combined mode. The targets have been selected from the 
near-infrared catalog of \citet{m06} for the innermost region (within $\sim$2.5 \arcmin\ from the cluster 
center) and from the 2MASS catalog \citep{skrutskie} for the external region.
The UVES fibers have been allocated in the circular corona between $\sim$20 \arcsec\ and $\sim$60 \arcsec\ 
from the cluster center.
The GIRAFFE fibers have been allocated in the surrounding field and only a few of them close to
the cluster center (note that the small angular size of the cluster, combined with the physical size 
of the magnetic button supporting the FLAMES fibers, prevents the allocation of 
more than $\sim$12 fibers in the central region).
The UVES targets were observed with the Red Arm 580 grating 
($\sim$4800-6800 \AA), while the GIRAFFE targets with
the HR11 ($\sim$5600-5800 \AA) and HR13 ($\sim$6100-6400 \AA) gratings. 
Both the UVES and GIRAFFE spectra were reduced using the dedicated ESO 
pipelines\footnote{http://www.eso.org/sci/software/pipelines/}.

\begin{figure}
\plotone{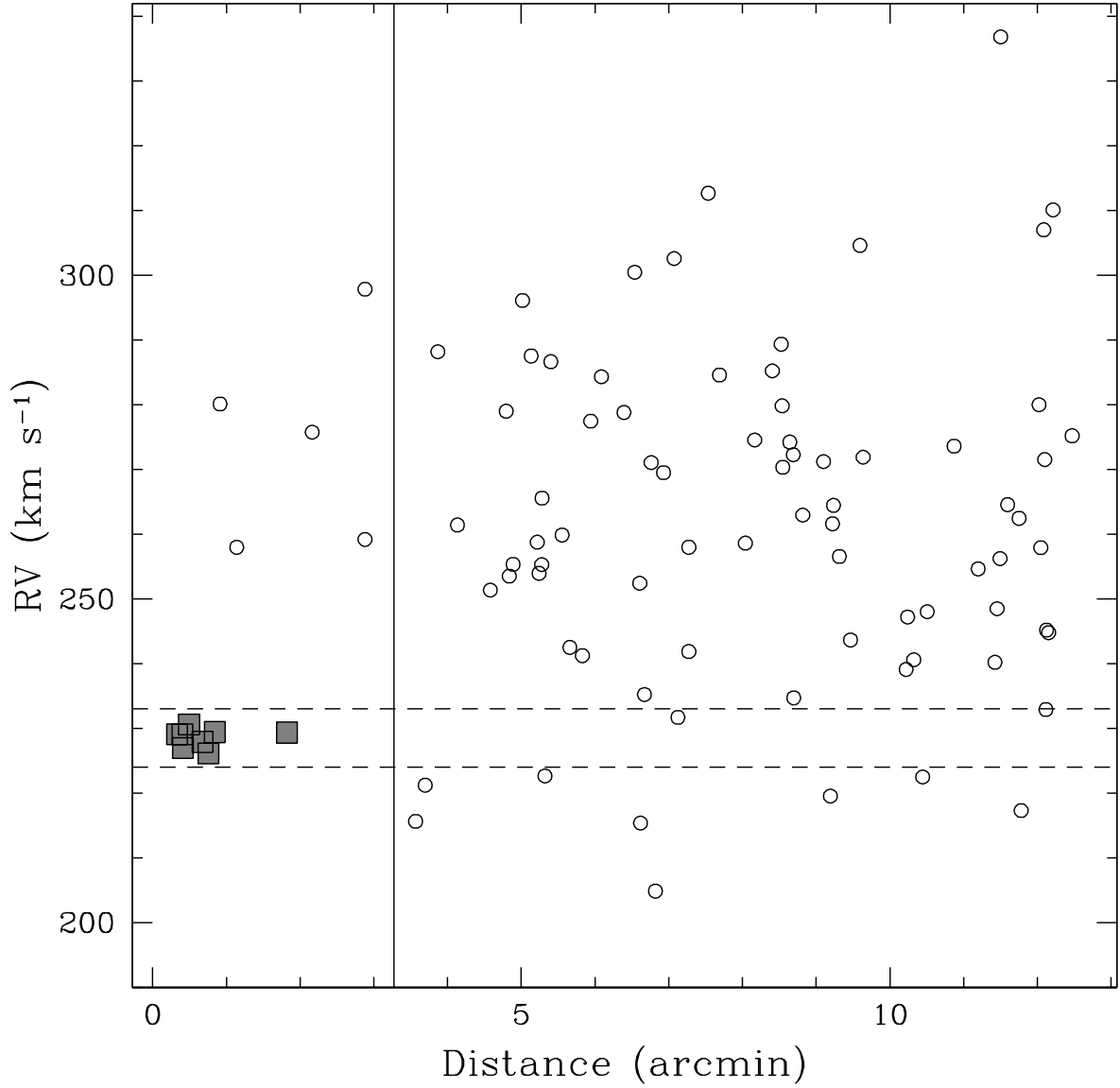}
\caption{Radial velocities as a function of the distance from the cluster 
center for the entire spectroscopic sample (gray squares are the cluster member stars 
and open circles the LMC field stars). 
The dashed horizontal lines indicate the $\pm$3$\sigma$ level from the average 
RV of the cluster. The vertical line indicates the tidal radius of NGC~1806 
\citep[$r_t$=~3.27 \arcmin\ , ][]{goud11}.}
\label{rvdis}
\end{figure}

Radial velocities (RVs) have been measured by means of DAOSPEC \citep{stetson}. 
Fig.~\ref{rvdis} shows the radial velocities of the 89 stars observed in the direction of 
NGC~1806 as a function of the distance from the cluster center. 
The RV distribution of the entire sample ranges from +204.9 to +336.8 \kms\ .
As can be seen in Fig.~\ref{rvdis} a clump of 8 stars is clearly visible at RV=+228.6$\pm$0.5 \kms\  
(gray squares). 
This value is fully compatible with RV=+225.0$\pm$5.6 \kms\  measured by \citet{ols91} 
for star members of NGC~1806. 
The 8 stars are located within 2 arcmin from the cluster center, well within 
the tidal radius \citep[$r_c$=~3.27 \arcmin\ ,][]{goud11} and within $\sim1.5\sigma$ from the mean 
cluster velocity.
The above considerations suggest that the 8 selected stars are {\sl bona fide} 
cluster members 
(this is also confirmed by the remarkable homogeneity in the derived 
iron abundance, see below).

Chemical abundances for a number of key elements (Fe, Ni, Mg, Al, Na, Si, Ca and Ti)
have been obtained for the entire sample of giant stars
by matching measured and theoretical equivalent widths (EWs)
using the package GALA \citep{m13g}\footnote{http://www.cosmic-lab.eu/gala/gala.php}, 
and  we used spectral synthesis only for O, in order to model the forbidden O line at 6300.3 \AA , 
affected by a close blending with a Ni transition. 
We refer the reader to \citet{m13_5694} for a detailed description of the adopted line list, 
model atmospheres and determination of the uncertainties.
EWs have been measured with the code DAOSPEC \citep{stetson}, run through the 
automatic wrapper 4DAO \citep{m13_4dao}\footnote{http://www.cosmic-lab.eu/4dao/4dao.php} 
that allows an analysis cascade of large sample of spectra and an easy, visual inspection 
of the Gaussian fit performed on each line of interest. 

Effective temperatures ($T_{eff}$) have been derived by means of the $(J-K_{s})_{0}-T_{eff}$ 
relation of \citet{ghb09} and adopting the color excess value of E(B-V)=~0.16 mag 
quoted by \citet{dirsch00}. Surface gravities have been derived with the Stefan-Boltzmann relation 
by assuming the photometric $T_{eff}$, the bolometric corrections of \citet{alonso99}, 
a distance modulus of $(m-M)_0$=~18.5 mag \citep{alves04} and a mass of 1.5$M_{\odot}$, 
according to a 1.75 Gyr-old isochrone \citep[the average value of the ages of the two MSTOs
quoted by][]{mackey08}\footnote{Even if the cluster is suspected to have a spread in age, 
the assumption of the average 
age to derive the stellar mass is not a crucial issue, because the proposed age spread of 
300 Myr corresponds to a mass difference of $\sim$0.08 $M_{\odot}$, with 
a negligible impact on the gravities ($\sim$0.02).}, metallicity Z=~0.004 and solar-scaled chemical mixture 
from the BaSTI data set \citep{pietr04}.
Finally, the microturbulent velocities have been derived spectroscopically, 
by erasing any trend between Fe~I abundance and line strength.   
Corrections for departure from LTE for the Na lines 
have been obtained by interpolating on the grid of \citet{gr99}.
Table 1 lists the abundance ratios measured for each member star of NGC~1806.
Details about the analysis and the discussion of the measured abundances 
for the entire sample will be reported in a forthcoming paper (A. Mucciarelli et al. 2014, in preparation).

-- {\it The photometric data set: }
We used a combination of HST Advanced Camera for Survey - Wide Field Camera (ACS/WFC)
and Wide Field Camera 3 - UVIS channel (WFC3/UVIS) data.
The ACS/WFC images (Prop ID: 10595; PI: Goudfrooij) were obtained with the F435W and F814W filters,
while the F336W band was used for the WFC3/UVIS (Prop ID: 12257, PI: Girardi) exposures.
The photometric reduction were performed by following the approach described in 
\citet{dalex14}.
The final catalog consists of stars detected in all bands. The resulting field of view
corresponds almost to the entire area covered by the WFC3 array ($\sim 160\arcsec \times
160\arcsec$). 
Instrumental magnitudes were reported to the VEGAMAG photometric system by using standard procedures and 
zero points
reported by \citet{sirianni05} for the ACS data set 
and in the WFC3 web page for the WFC3 sample.
Instrumental coordinates were roto-translated to the absolute $(\alpha, \delta)$ coordinate system
using the stars in common with the 2MASS catalog.
Figure~\ref{cmd} shows the ($m_{F814W}$; $m_{F336W}-m_{F814W}$) CMD of NGC~1806 for the stars lying within
about $30\arcsec$ from the cluster center.

\section{Searching for chemical inhomogeneities in NGC~1806}

\begin{figure}
\plotone{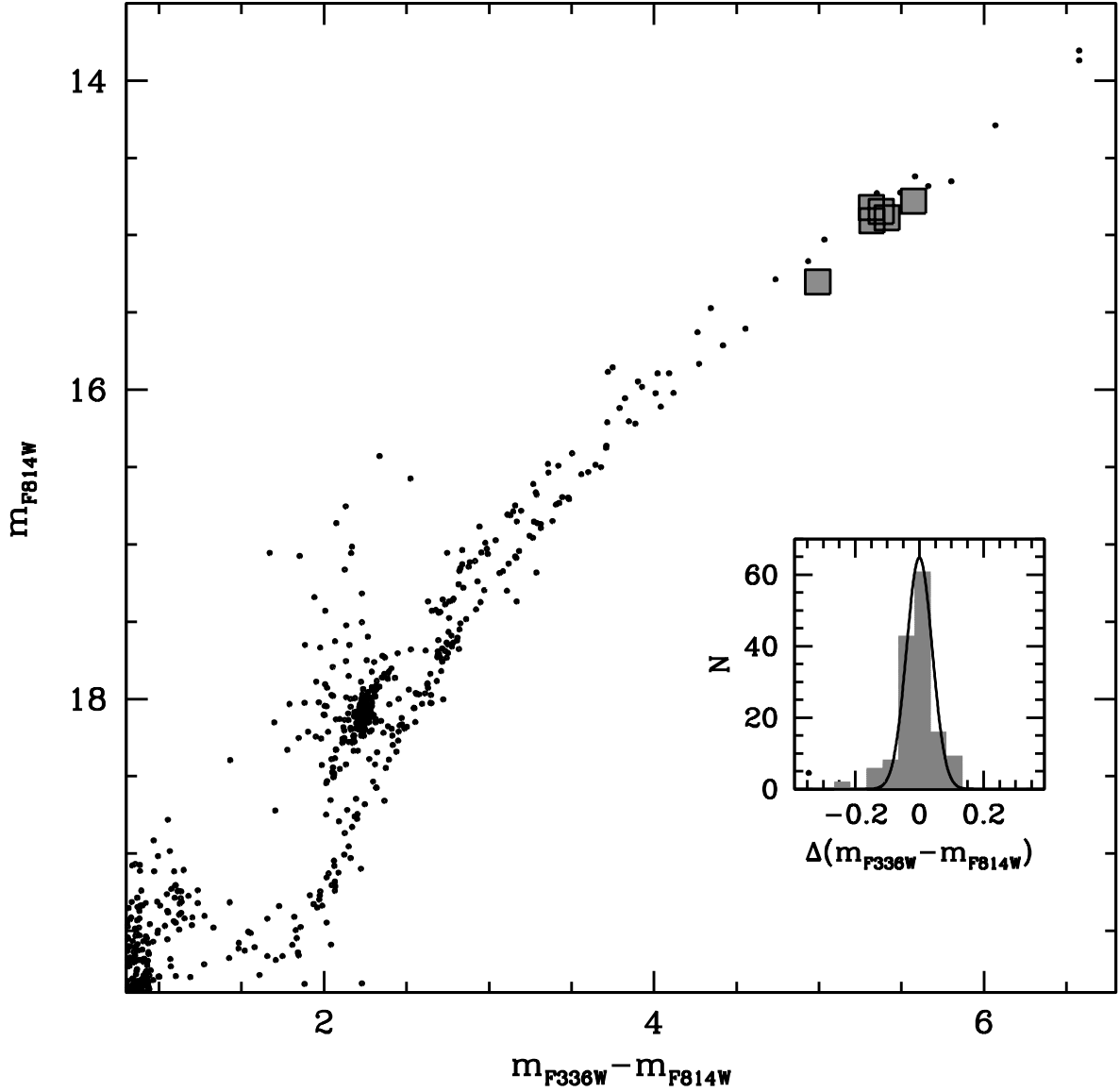}
\caption{($m_{F814W}$; $m_{F336W}-m_{F814W}$) color-magnitude diagram obtained with HST data 
for the innermost 30 \arcsec\ from the cluster center. 
Gray squares are the spectroscopic targets  lying in HST field of view.
The inset shows the histogram of the ($m_{F336W}-m_{F814W}$) color residuals calculated 
with respect to the RGB mean ridge line, with over-imposed the distribution of the photometric 
uncertainties.}
\label{cmd}
\end{figure}

\begin{figure}
\plotone{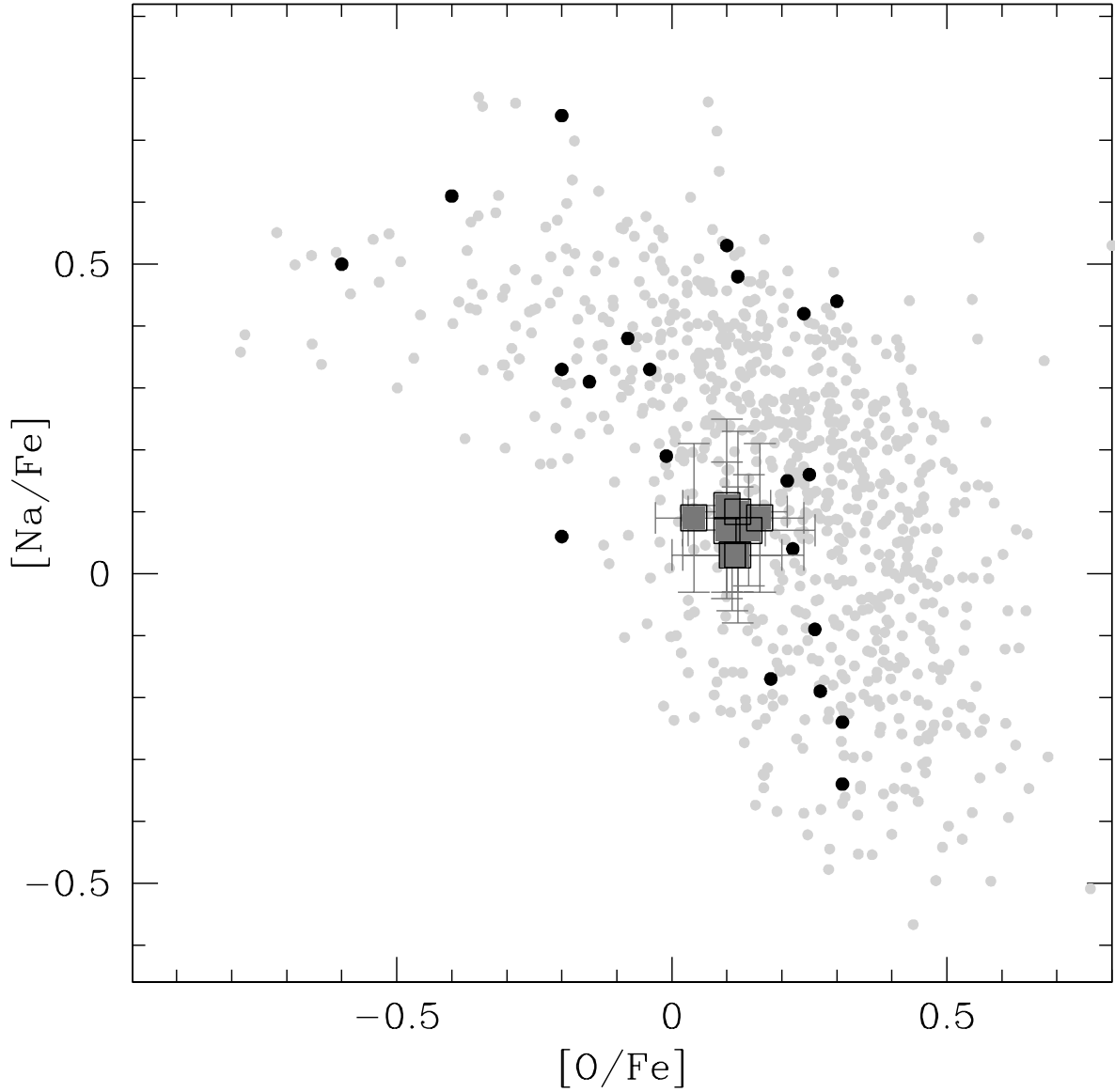}
\caption{[Na/Fe] as a function of [O/Fe] for the 8 members of NGC~1806 
(dark gray squares), compared to the stars in old Milky Way \citep[light gray points,][]{carretta09} 
and LMC \citep[black points,][]{m09} GCs. }
\label{nao}
\end{figure}

\begin{figure}
\plotone{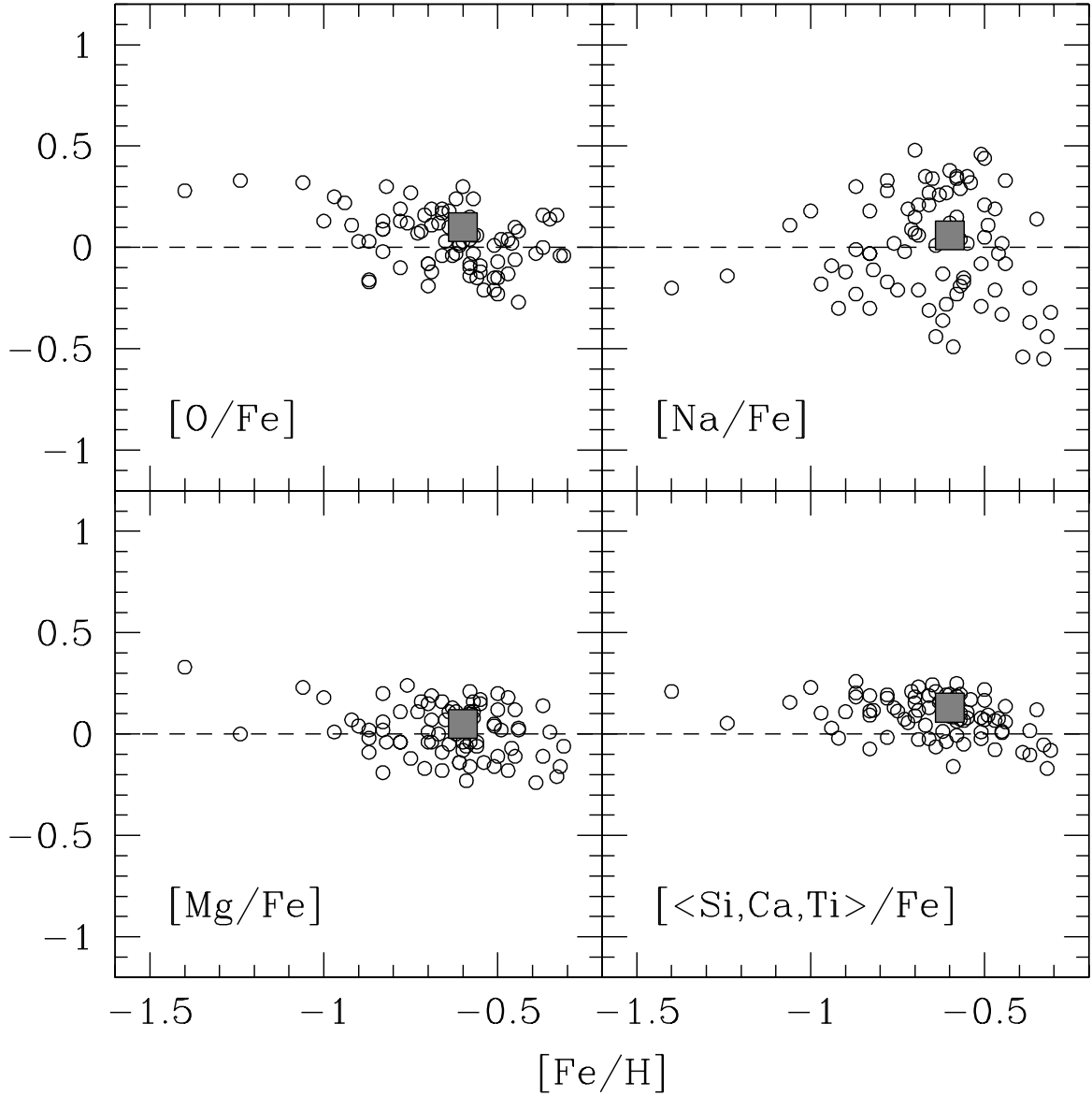}
\caption{Comparison between the average values of the [O/Fe], [Na/Fe], [Mg/Fe] and 
[$<$Si,Ca,Ti$>$/Fe] abundance ratios for NGC~1806 (gray squares) and for the surrounding field stars 
(open circles).}
\label{fld}
\end{figure}

For the 8 stars considered to be {\sl bona fide} members of 
NGC~1806, we obtained the following results:
\begin{enumerate}
\item 
No evidence of star-to-star variations in terms of iron and light elements 
has been found. The average iron abundance turns out to be 
[Fe/H]=--0.60$\pm0.01$ dex ($\sigma_{obs}$=~0.02 dex).
The same level of homogeneity has been obtained 
for the light elements (O, Na, Al, Mg), which exhibit intrinsic spreads 
in all old, massive GCs,  both in our Galaxy \citep{carretta09} and 
in the LMC \citep{m09}. 
We carefully checked for the presence of intrinsic scatter by using the Maximum Likelihood 
algorithm described in \citet{m12_2419}, which calculates the mean and the intrinsic spread 
($\sigma_{int}$, with their 
corresponding uncertainties) of a given abundance ratio by taking into account the uncertainties 
of each individual star. 
For all the measured abundance ratios, the target stars formally have $\sigma_{int}$=~0.0,
pointing out a strict homogeneity in the chemical content of NGC~1806. 
The lack of any significant light element spread can be appreciated in Fig.~\ref{nao}, where the 
individual stars of NGC~1806 (dark gray squares) are compared to the stars observed in old Galactic 
and LMC GCs (\citet{carretta09} and \citet{m09}, respectively) in the [O/Fe]--[Na/Fe] plane.

\item 
Fig.~\ref{fld} shows the average abundance ratios measured for the 
8 cluster members and the abundance ratios for the surrounding LMC field stars, both for three elements 
(O, Na, Mg) involved in the well known chemical anomalies and for 
[$<$Si,Ca,Ti$>$/Fe] which usually shows no star-to-star variations.
Oxygen and magnesium abundances nicely agree with those measured for other $\alpha$-elements 
(Si, Ca, Ti), at variance with what was found in old GCs where the stars with light element
anomalies show  [O/Fe] and [Mg/Fe] abundance ratios systematically lower than the other 
$\alpha$-elements. 
In addition,  the abundance ratios measured in NGC~1806 well resemble those derived in the 
surrounding field stars (which show solar-scaled abundances, on average).
Generally, in an old GC showing light-elements spread, stars with chemistry similar 
to that of the field are considered to be first generation stars, while those with 
significant light element differences are labeled as second generation stars.
In the case of NGC~1806, the good agreement between cluster and field stars
suggests that all the observed cluster members formed in the first episode of star formation.

\item At odds with what is typically found in old GCs
properly observed in the U band  \citep[or similar filters, see e.g. the cases of M4 by \citet{marino08} and 
NGC~6362 by][]{dalex14}, 
the ($m_{F814W}$; $m_{F336W}-m_{F814W}$) CMD of NGC~1806 does not exhibit any appreciable broadening 
or splitting of the RGB (see Fig.~\ref{cmd}). 
In this plane possible RGB splittings are essentially driven by
the intrinsic variations of C and N abundances, because the spectral region covered by the 
F336W filter (centered around $\sim$3360 $\AA$) is dominated by strong CN and NH molecular 
bands \citep{sbordone11}. The distribution of the ($m_{F336W}-m_{F814W}$) color 
residuals calculated with respect to the RGB mean ridge line (see the inset in Fig.~\ref{cmd})
is fully compatible with the photometric uncertainties.
This result suggests that no variations in the C and N abundances occur in the 
stellar population of NGC~1806.  

\end{enumerate}
Although the size of the presented spectroscopic sample is admittedly small, we note 
that chemical anomalies as large as those observed in the Galactic GCs have been 
detected in old LMC GCs \citep[namely NGC~1786, NGC~2210 and NGC~2257,][]{m09} 
on the basis of samples of similar size.
On the other hand, in old GCs the second generation stars
numerically dominate the cluster population 
\citep[accounting for $\sim$50\%--70\% of the entire stellar content;][]{carretta09}. 
Thus, if two distinct stellar populations were present in NGC~1806, we would have 
expected to observe $\sim$4--6 stars with significant 
light element abundance differences. 
Also, the remarkable chemical similarity between the cluster and the surrounding field stars 
rules out the possibility that all the observed stars are second generation stars.

\section{Discussion and conclusions}

The presence of bimodal or extended MSTOs observed in some Magellanic intermediate-age clusters  
and the chemical anomalies in light element abundances observed in old stellar 
clusters
have been suggested to have the same origin \citep{goud09,keller11}: 
they could be the fingerprint of self-enrichment processes operated by intermediate-mass 
AGBs and/or fast-rotating massive stars over a short timescale at the epoch of the cluster 
formation.
Hence, stars in clusters with extended MSTOs 
should exhibit significant light element abundance variations. 
The analysis of the chemical and photometric properties of the bimodal MSTO cluster 
NGC~1806 presented in this Letter instead, indicates a  high level of chemical 
homogeneity in the cluster population, drawing a simple conclusion: 
{\sl NGC~1806 does not show major evidence of multiple stellar populations 
as observed in old GCs}
\citep[included the old LMC GCs that exhibit chemical anomalies 
like those of the Galactic ones,][]{m09}.
Also, we recall that NGC~1806 has a present-day mass \citep[M=1.07$\cdot10^{5}M_{\odot}$,][]{goud11b} 
comparable with those of the Galactic/LMC old clusters where the chemical anomalies 
have been observed.

Hence, there is no evidence that 
NGC~1806 underwent any self-enrichment process. This suggests univocally that bimodal MSTO 
and light element chemical anomalies cannot be explained within the same framework. 
Moreover, a significant broadening of the MSTO has been observed only in a relatively 
small range of cluster ages, 
between 1 and 2 Gyr \citep[see Figure~1 in][]{keller11}, while it is not visible in younger 
clusters at odds with what would be expected if it was a common feature of the cluster formation 
process. Indeed, for example, the young, massive LMC cluster NGC~1866 
has been found to harbor a mono-age \citep{bastian13b} and
mono-metallic \citep{m11} stellar population.

In light of this finding, other solutions (invoking or not 
an intrinsic age spread but always preserving the homogeneity in the chemical abundances) 
should be envisaged to explain this class of globulars 
\citep[see e.g. the extended discussions provided by][and \citealt{keller11}]{santiago02,bekki}.
As suggested by \citet{bekki}, new episodes of 
star formation could be triggered by the merging or the interaction between an already existing 
cluster (corresponding to the faintest MSTO) and a giant molecular cloud. 
While no  explicit predictions  are provided for the chemical content of a GC
formed through this mechanism,  the authors do not exclude that the pollution by AGB stars 
could occur and that the stars formed from the second burst of star formation could display 
the same kind of chemical anomalies in light elements observed in old GCs. 
However, this could be in disagreement with our findings in NGC~1806.

Another possibility is that bimodal MSTO GCs are the result of a merging event between 
the components of a binary/multiple cluster system, formed by the collapse of the 
same molecular cloud but with an initial difference in their ages. 
This would naturally account for chemical homogeneity in clusters with 
different ages.
The LMC and SMC harbor a significant population of candidate binary/multiple clusters 
\citep[see][and references therein]{m12_bin}. 
This scenario could explain only the bimodal MSTOs, but cannot account for clusters
with an 
extended MSTO without discrete sub-structures.

Other mechanisms, not requiring an age spread, could be the solution and they 
should be better investigated. 
\citet{bastian09} first proposed that the effect of rotation in the evolution of MS stars 
with M$>$1.2$M_{\odot}$ can account for the observed MSTO:
for a coeval population, fast rotating MS stars evolve to lower $T_{eff}$, with respect to non-rotating stars, 
thus mimicking a redder MSTO. 
However, by including the effect of the rotation in the calculation of the stellar lifetimes 
\citep[an effect neglected by ][]{bastian09} 
\citet{girardi11} conclude that rotating stars are slightly bluer and brighter 
than non-rotating ones, but this effect is not sufficient to mimic extended MSTOs
like those observed in the Magellanic GCs. On the other hand, 
the recent investigation by \citet{yang13} leads to the opposite result. 
Their calculations show that rotating stars are colder and evolve faster than 
non-rotating stars.
In particular, \citet{yang13} emphasize that the efficiency of rotational induced 
mixing (a mechanism that is still debated) is a crucial parameter to reproduce the observed 
MSTO. Thus, they conclude that rotation can, in principle, lead to a dual or broad MSTO. 
Further investigations of this process and any alternative scenarios are needed and urged 
to finally understand the origin of the observed MSTO morphologies.

\acknowledgements  
We thank the anonymous referee for useful comments.
This research is part of the project COSMIC-LAB (Web site: http://www.cosmic-lab.eu) 
funded by the European Research Council (under contract ERC-2010-AdG-267675).

{}

\begin{landscape}
\begin{deluxetable}{lccccccccc}
\tablecolumns{10} 
\tiny
\tablewidth{0pc}  
\tablecaption{}
\tablehead{ 
\colhead{Star} &   RA & 
Dec & J & K & RV & $T_{eff}$ & log~g & $v_{turb}$ &
[Fe/H] \\
  &   (J2000)  &  (J2000)  &  & &(\kms\ ) & (K) & & (\kms\ )  & (dex)  }
\startdata 
\hline
 NGC~1806-25  &  75.5685190   &  -67.9808206	 &  13.68   & 12.65  &  230.6$\pm$0.   & 3920  &  0.7  & 1.60  & --0.62$\pm$0.04	\\    
 NGC~1806-27  &  75.5639398   &  -67.9835530	 &  13.70   & 12.69  &  229.1$\pm$0.   & 3990  &  0.7  & 1.60  & --0.58$\pm$0.04	\\    
 NGC~1806-29  &  75.5152945   &  -67.9793630	 &  13.74   & 12.74  &  229.4$\pm$0.   & 4000  &  0.7  & 1.70  & --0.62$\pm$0.04	\\    
 NGC~1806-30  &  75.5473719   &  -67.9918053	 &  13.74   & 12.79  &  227.0$\pm$0.   & 4120  &  0.8  & 1.60  & --0.58$\pm$0.05	\\    
 NGC~1806-32  &  75.5162281   &  -67.9868209	 &  13.76   & 12.80  &  226.1$\pm$0.   & 4090  &  0.8  & 1.50  & --0.59$\pm$0.05	\\    
 NGC~1806-40  &  75.5323035   &  -67.9868328	 &  14.28   & 13.35  &  229.0$\pm$0.   & 4150  &  1.0  & 1.60  & --0.58$\pm$0.04     \\ 	
 NGC~1806-33  &  75.5638527   &  -67.9949059	 &  13.85   & 12.86  &  227.9$\pm$0.   & 4040  &  0.8  & 1.60  & --0.61$\pm$0.11     \\        
 NGC~1806-39  &  75.5908674   &  -67.9588433	 &  14.13   & 13.22  &  229.3$\pm$0.   & 4220  &  1.0  & 1.60  & --0.60$\pm$0.10       \\    
\hline
\hline
\colhead{Star} & [O/Fe]  &  [Na/Fe]  &  [Mg/Fe]  &  [Al/Fe]  &  [Si/Fe]  &  [Ca/Fe]  &  [Ti/Fe]  &  [Ni/Fe] & \\
 & (dex)  &   (dex)   &  (dex)    & (dex) & (dex) &(dex) & (dex) & (dex) & \\ 
\hline 
  NGC~1806-25	& +0.10$\pm$0.08       &    +0.11$\pm$0.14  &  0.06$\pm$0.07   &+0.09$\pm$0.12    &    +0.15$\pm$0.10	& +0.02$\pm$0.15    &	+0.09$\pm$0.18   &  --0.08$\pm$0.03   & 	  \\ 
  NGC~1806-27	& +0.04$\pm$0.07       &    +0.09$\pm$0.12  &  0.02$\pm$0.06   &+0.14$\pm$0.09    &    +0.08$\pm$0.11	& +0.09$\pm$0.12    &	+0.13$\pm$0.16   &  --0.13$\pm$0.04   & 	  \\ 
  NGC~1806-29	& +0.16$\pm$0.08       &    +0.09$\pm$0.12  &--0.01$\pm$0.07   &+0.11$\pm$0.09    &    +0.11$\pm$0.11	& +0.14$\pm$0.13    &	+0.17$\pm$0.16   &  --0.15$\pm$0.04   & 	  \\ 
  NGC~1806-30	& +0.12$\pm$0.09       &    +0.10$\pm$0.13  &  0.06$\pm$0.08   &+0.19$\pm$0.08    &    +0.03$\pm$0.11	& +0.16$\pm$0.11    &	+0.25$\pm$0.15   &  --0.17$\pm$0.05   & 	  \\ 
  NGC~1806-32	& +0.10$\pm$0.07       &    +0.07$\pm$0.11  &  0.11$\pm$0.07   &+0.19$\pm$0.09    &    +0.02$\pm$0.10	& +0.13$\pm$0.11    &	+0.20$\pm$0.15   &  --0.16$\pm$0.04   & 	  \\ 
  NGC~1806-40	& +0.11$\pm$0.09       &    +0.03$\pm$0.09  &  0.06$\pm$0.08   &+0.13$\pm$0.08    &    +0.07$\pm$0.11	& +0.14$\pm$0.10    &	+0.12$\pm$0.15   &  --0.14$\pm$0.04   &  \\ 
  NGC~1806-33	& +0.12$\pm$0.12       &    +0.03$\pm$0.11  &  0.12$\pm$0.13   &    --- 	  &    +0.16$\pm$0.13	& +0.16$\pm$0.12    &	+0.15$\pm$0.17   &  --0.12$\pm$0.08   &   \\ 
  NGC~1806-39	& +0.14$\pm$0.12       &    +0.07$\pm$0.09  &  0.09$\pm$0.10   &    --- 	  &    +0.14$\pm$0.12	& +0.17$\pm$0.11    &	+0.14$\pm$0.14   &  --0.08$\pm$0.08   &   \\ 
 \hline
\enddata 
\tablecomments{$~~~~~$Coordinates and magnitudes are 
from \citet{m06}. The abundance uncertainties  include the internal error and 
that due to the atmospheric parameters.}
\end{deluxetable}
\end{landscape}

\begin{thebibliography}{}

\bibitem[Alonso et al.(1999)]{alonso99}
Alonso, A., Arribas, S., \& Martinez-Roger, C., 1999, A\&As, 140, 261

\bibitem[Alves(2004)]{alves04}
Alves, D. R., 2004, New Astro. Rev., 48, 659

\bibitem[Bastian \& de Mink(2009)]{bastian09}
Bastian, N., \& de Mink, S. E., 2009, MNRAS, 398, L11

\bibitem[Bastian et al.(2013)]{bastian13a}
Bastian, N., Lamers, H. J. G. L. M., de Mink, S. E., Longmore, S. N., 
Goodwin, S. P., \& Gieles, M., 2013, MNRAS, 436, 2398

\bibitem[Bastian \& Silva-Villa(2013)]{bastian13b}
Bastian, N., \& Silva-Villa, E., 2013, MNRAS, 431, 122

\bibitem[Bekki \& Mackey(2009)]{bekki}
Bekki, K., \& Mackey, A. D., 2009, MNRAS, 394, 124


\bibitem[Carretta et al.(2009a)]{carretta09}  
Carretta, E., et al., 2009, A\&A, 505, 117

\bibitem[Carretta et al.(2009b)]{carretta09b}  
Carretta, E., Bragaglia, A., Gratton, R., D'Orazi, V., \& Lucatello, S., 2009, A\&A, 505, 139

\bibitem[Dalessandro et al.(2014)]{dalex14}
Dalessandro, E., Massari, D., Bellazzini, M., Miocchi P., Mucciarelli, A., 
Salaris, M., Cassisi, S., Ferraro, F. R., \& Lanzoni, B., 2014, ApJ, 791L, 4

\bibitem[Decressin et al.(2010)]{decressin10}
Decressin, T., Baumgardt, H., Charbonnel, C., \& Kroupa, P., 2010, A\&A, 516, 73

\bibitem[D'Ercole et al.(2008)]{dercole08}
D'Ercole, A., Vesperini, E., D'Antona, F., McMillan, S. L. W., \& Recchi, S., 
2008, MNRAS, 391, 825

\bibitem[Dirsch et al.(2000)]{dirsch00}
Dirsch, B., Richtler, T., Gieren, W. P., \& Hilker, M., 2000, A\&A, 360, 133

\bibitem[Glatt et al.(2008)]{glatt08}
Glatt, K., et al., 2008, AJ, 136, 1703

\bibitem[Girardi et al.(2011)]{girardi11}
Girardi, L.,  Eggenberger, P., \& Miglio, A., 2011, MNRAS, 412, L103

\bibitem[Gonzalez Hernandez \& Bonifacio(2009)]{ghb09}
Gonzalez Hernandez, J. I., \& Bonifacio, P., 2009, A\&A, 497, 497

\bibitem[Goudfrooij et al.(2009)]{goud09}
Goudfrooij, P, Puzia, T. H., Kozhurina-Platais, V., \& Chandar, R., 2009, AJ, 137, 4988

\bibitem[Goudfrooij et al.(2011a)]{goud11}
Goudfrooij, P, Puzia, T. H., Kozhurina-Platais, V., \& Chandar, R., 2011, ApJ, 737, 3

\bibitem[Goudfrooij et al.(2011b)]{goud11b}
Goudfrooij, P, Puzia, T. H., Chandar, R., \& Kozhurina-Platais, V., 2011, ApJ, 737, 4

\bibitem[Gratton et al.(1999)]{gr99}
Gratton, R. G., Carretta, E., Eriksson, K. \& Gustafsson, B., 1999, A\&A, 350, 955

\bibitem[Gratton et al.(2012)]{gratton12}
Gratton, R. G., Carretta, E., \& Bragaglia, A., 2012, A\&ARv, 20, 50
\
\bibitem[Keller et al.(2011)]{keller11}
Keller, S. C., Mackey, A. D., \& Da Costa, G. S., 2011, ApJ, 731, 22

\bibitem[Mackey \& Broby Nielsen(2007)]{mackey07}
Mackey, A. D., \& Broby Nielsen, P., 2007, MNRAS, 379, 151

\bibitem[Mackey et al.(2008)]{mackey08}
Mackey, A. D., Broby Nielsen, P., Ferguson, A. M. N., \& Richardson, J. C., 
2008, ApJ, 681L, 17

\bibitem[Marino et al.(2008)]{marino08}
Marino, A. F., Villanova, S., Piotto, G., Milone, A. P., Momany, Y.,  Bedin, L. R., \& Medling, A. M., 
2008, A\&A, 490, 625

\bibitem[Milone et al.(2009)]{milone09}
Milone, A. P., Bedin, L. R., Piotto, G., \& Anderson, J., 2009, A\&A, 497, 755

\bibitem[Mucciarelli et al.(2006)]{m06}
Mucciarelli, A., Origlia, L., Ferraro, F. R., Maraston, C., \& Testa, V., 2006, ApJ, 646, 939

\bibitem[Mucciarelli et al.(2007)]{m07_1978}
Mucciarelli, A., Ferraro, F. R., Origlia, L., \& Fusi Pecci, F., 2007, AJ, 133, 2053



\bibitem[Mucciarelli et al.(2009)]{m09}
Mucciarelli, A., Origlia, L., Ferraro, F. R., \& Pancino, E., 2009, ApJ, 695L, 134

\bibitem[Mucciarelli et al.(2011)]{m11}
Mucciarelli, A., Cristallo, S., Brocato, E., Pasquini, L., Straniero, O., Caffau, E., 
Raimondo, G., Kaufer, A., Musella, I., Ripepi, V., Romaniello, M., \& Walker, A. R., 
2011, MNRAS, 413, 837

\bibitem[Mucciarelli et al.(2012a)]{m12_2419}
Mucciarelli, A., Bellazzini, M., Ibata, R., Merle, T., Chapman, S. C., 
Dalessandro, E., \& Sollima, A., 2012, MNRAS, 426, 2889

\bibitem[Mucciarelli et al.(2012b)]{m12_bin}
Mucciarelli, A., Origlia, L., Ferraro, F. R., Bellazzini, M., \& Lanzoni, B., 2012, ApJ, 746L, 19

\bibitem[Mucciarelli et al.(2013a)]{m13g}
Mucciarelli, A., Pancino, E., Lovisi, L., Ferraro, F. R., \& Lapenna, E., 2013, ApJ, 766, 78

\bibitem[Mucciarelli et al.(2013b)]{m13_5694}
Mucciarelli, A., Bellazzini, M., Catelan, M., Dalessandro, E., Amigo, P., Correnti, M., 
Cort\'es, C.\&  D'Orazi, V, 2013, MNRAS, 435, 3667

\bibitem[Mucciarelli(2013c)]{m13_4dao}
Mucciarelli, A., 2013, arXiv1311.1403


\bibitem[Olszewski et al.(1991)]{ols91}
Olszewski, E. W., Schommer, R. A., Suntzeff, N. B., \& Harris, H. C., 1991, AJ, 101, 515

\bibitem[Pasquini et al.(2000)]{pasquini}
Pasquini, L., et al., 2000, SPIE, 4008, 129

\bibitem[Pietrinferni et al.(2004)]{pietr04}
Pietrinferni, A., Cassisi, S., Salaris, M., \& Castelli, F., 2004, ApJ, 612, 168

\bibitem[Rubele et al.(2013)]{rubele13}
Rubele, S., Girardi, L., Kozhurina-Platais, V., Kerber, L., Goudfrooij, P., 
Bressan, A., \& Marigo, P., 2013, MNRAS, 430, 2774

\bibitem[Santiago et al.(2002)]{santiago02}
Santiago, B., Kerber, L., Castro, R., \& de Grijs, R., 2002 , MNRAS, 336, 139

\bibitem[Sbordone et al.(2011)]{sbordone11}
Sbordone, L., Salaris, M., Weiss, A., \& Cassisi, S., 2011, A\&A, 534, 9

\bibitem[Skrutskie et al.(2006)]{skrutskie}
Skrutskie, M. F., et al., 2006, AJ, 131, 1163

\bibitem[Sirianni et al.(2005)]{sirianni05}
Sirianni, M., et al., 2005, PASP, 117, 1049

\bibitem[Stetson \& Pancino(2008)]{stetson}
Stetson, P. \& Pancino, E., 2008, PASP, 120, 1332

\bibitem[Valcarce \& Catelan(2011)]{valcarce11}
Valcarce, A. A. R., \& Catelan, M., 2011, A\&A, 533, 120

\bibitem[Ventura et al.(2001)]{ventura01}
Ventura, P., D'Antona, F., Mazzitelli, I., \& Gratton, R., 
2001, ApJ, 550, 65

\bibitem[Yang et al.(2011)]{yang11}
Yang, W., Meng, X., Bi, S., Tian, Z., Li, T., \& Liu, K., 2011, ApJ, 731L, 37

\bibitem[Yang et al.(2013)]{yang13}
Yang, W., Bi, S., Meng, X., \& Liu, K., 2013, ApJ, 776, 112

\end{thebibliography}
\end{document}